\documentclass[12pt]{article}
\def\ca{\c{c}\~{a}}

\usepackage{epsfig}
\usepackage{cite}
\long\def\symbolfootnote[#1]#2{\begingroup\def\thefootnote{\fnsymbol{footnote}}\footnote[#1]{#2}\endgroup}

\begin{document}

{\centerline{\bf Pion quadrupole polarizabilities in the Nambu--Jona-Lasinio 
model\symbolfootnote[1]{We are very grateful to the organizers of the ``Mini-Workshop Bled 2009: Problems in multi-quark states'', for the kind invitation to present this work.    
This research is supported by the Polish Ministry of Science and Higher
Education, grants N202~034~32/0918 and N202~249235, by Funda\ca o para a 
Ci\^encia e Tecnologia, grants FEDER, OE, POCI 2010, CERN/FP/83510/2008, and 
by the European Community-Research 
Infrastructure Integrating Activity “Study of Strongly Interacting Matter” 
(Grant Agreement 227431) under the Seventh Framework Programme of EU.}}}

\vspace{0.5cm}

{\centerline{ Brigitte Hiller$^1$, Wojciech Broniowski$^{2,3}$, Alexander A. Osipov$^{1,4}$, Alex H. Blin$^1$}}
\vspace{0.5cm}

\noindent{\it{$^1$Centro de F\'{\i}sica Computacional, Departamento de 
F\'{\i}sica da Universidade de Coimbra, 3004-516 Coimbra, Portugal}

\noindent{$^2$The H.~Niewodnicza\'nski Institute of Nuclear Physics, Polish 
Academy of Sciences, PL-31342 Cracow, Poland}

\noindent{$^3$Institute of Physics, Jan Kochanowski University, PL-25406~Kielce, Poland}

\noindent{$^4$Dzhelepov Laboratory of Nuclear Problems, JINR, 141980 Dubna, Russia}}
\vspace{0.5cm}

\begin{abstract}
We present the results for the pion electromagnetic quadrupole polarizabilities,
calculated within the Nambu--Jona-Lasinio model. We obtain the sign and magnitude in agreement
with the respective experimental analysis based on the Dispersion Sum Rules.
At the same time the dipole polarizabilities are well reproduced. Comparison is also made with the results from  
the Chiral Perturbation Theory.
\end{abstract}

The neutral and charged pion  dipole and quadrupole polarizabilities have been recently analyzed using the Dispersion Relations (DR) and 
the Dispersion Sum Rules (DSR) \cite{Fil'kov:2005,Fil'kov:2006}, as displayed in
Tables \ref{tab:N} and \ref{tab:V}, together with the results of the Chiral Perturbation Theory ($\chi$PT) \cite{Gasser:2005,Gasser:2006,Bellucci:1994}. The first row shows our results based on the  Nambu--Jona-Lasinio model (NJL) \cite{Nambu:1961} model; for that purpose we have extended~\cite{Hiller:2009} the study of Ref.~\cite{Bajc:1996}, where the dipole polarizabilities have been calculated, to the quadrupole case. We refer to these papers for details. 

\begin{figure}[tb]
\begin{center}
\includegraphics[width=0.28\textwidth]{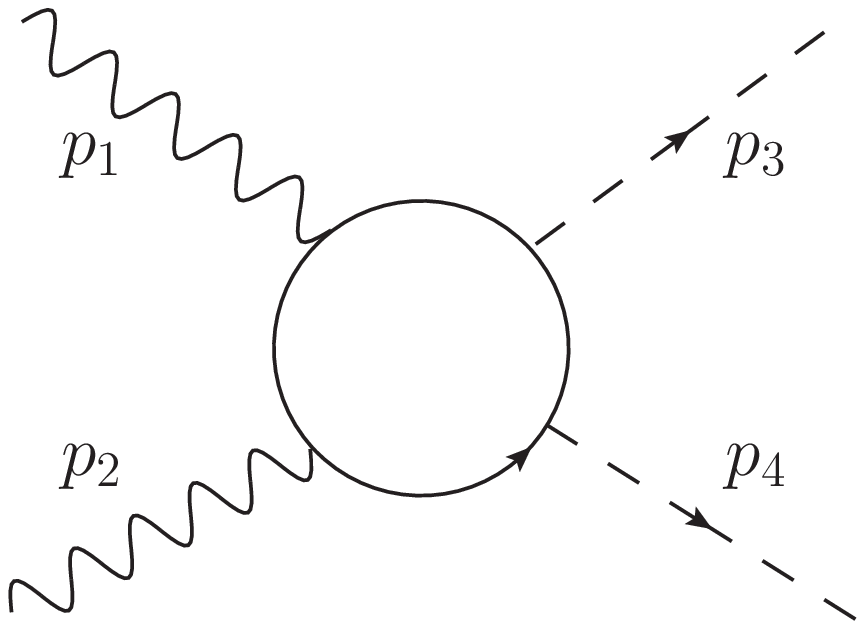}
\includegraphics[width=0.35\textwidth]{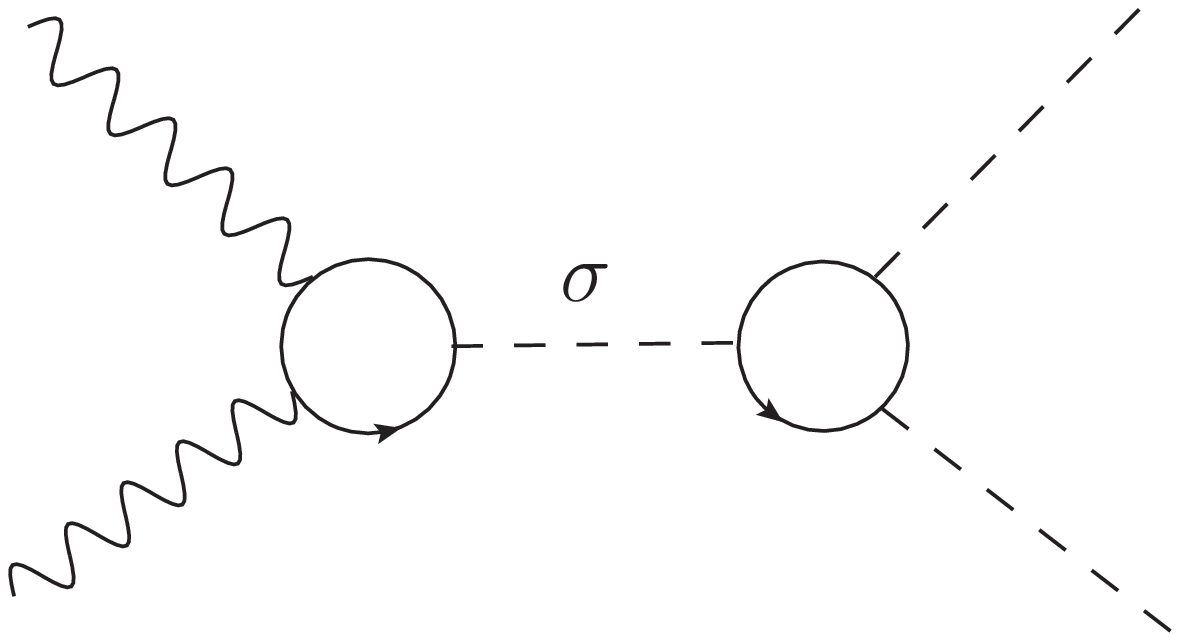}
\includegraphics[width=0.29\textwidth]{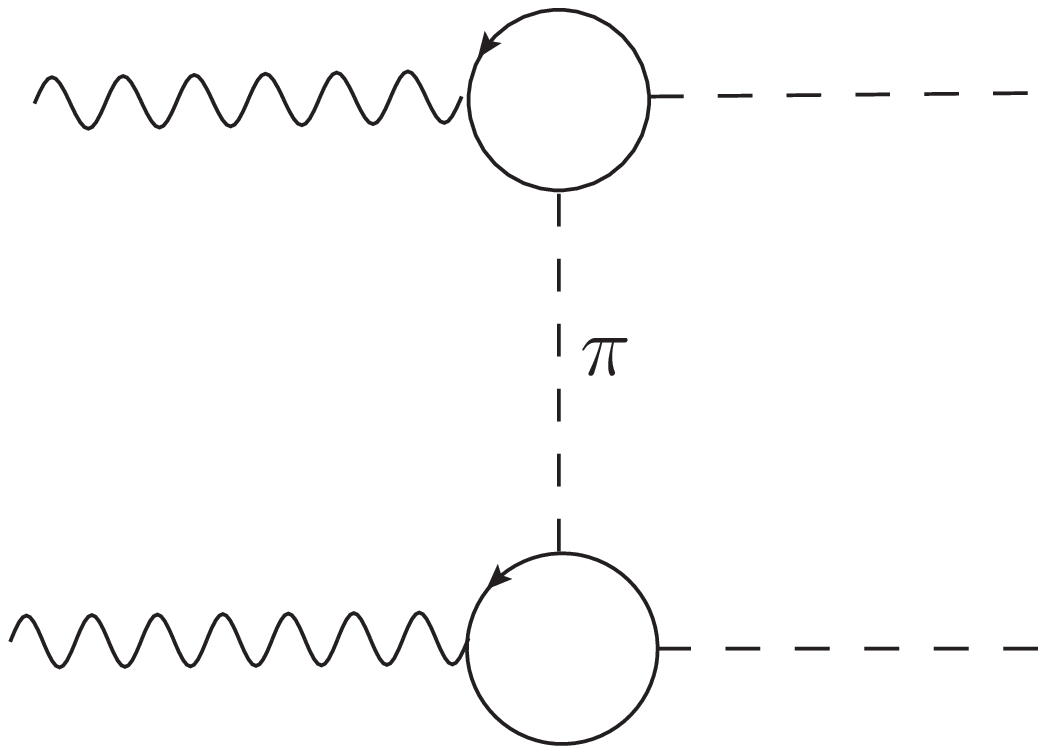}
\caption{Leading-$N_c$ quark-loop diagrams for the $\gamma\gamma\to\pi\pi$ 
amplitude. The crossed terms are not displayed. 
\label{fig:graphs}}
\end{center}
\end{figure}

Our leading-$N_c$ calculations are done according to the Feynman 
diagrams of Fig.~\ref{fig:graphs}. 
The 
amplitude is a  function of the Mandelstam variables related to the 
$\gamma(p_1,\epsilon_1)+\gamma(p_2,\epsilon_2)\to\pi^a(p_3)+\pi^b(p_4)$ 
reaction for the on-shell pions and photons,  
\begin{equation}
\label{amplitudes}
   T(p_1,p_2,p_3)=e^2\epsilon_1^\mu\epsilon_2^\nu T_{\mu\nu}, \quad
   T_{\mu\nu}=A(s,t,u) {\cal L}_1^{\mu\nu} +B(s,t,u) {\cal L}_2^{\mu\nu},
\end{equation}
with the Lorentz tensors   
\begin{equation}
\label{invariants}
   {\cal L}_1^{\mu\nu}=p_2^\mu p_1^\nu-\frac{1}{2}s g^{\mu\nu}, 
%   \nonumber \\ 
\hspace{1cm}
   {\cal L}_2^{\mu\nu}=-\left(\frac{1}{2}u_1t_1 g^{\mu\nu} 
   +t_1 p_2^\mu p_3^\nu +u_1 p_3^\mu p_1^\nu +s p_3^\mu p_3^\nu\right).
\end{equation}
Terms that vanish upon the conditions
$\epsilon_1\cdot p_1=\epsilon_2 \cdot p_2=0$ are omitted and the notation  
$\xi_1=\xi -m_\pi^2$, $\xi=s,t,u$ is used. The scalar quantities $A$ and $B$ enter 
the amplitudes $H_{++}=-(A+m_\pi^2 B)$ and $H_{+-}=({u_1t_1}/{s}-
m_\pi^2)B$ for the equal-helicity and helicity-flipped photons.  The dipole, $\alpha_1^i,\beta_1^i$, and quadrupole, 
$\alpha_2^i,\beta_2^i$, polarizabilities are obtained in the $t$-channel and extracted 
from the first two coefficients of the Taylor expansion of 
the amplitudes $\alpha A^i(s,t,u)/(2 m_\pi)$ and  $-\alpha m_\pi B^i(s,t,u)$ around $s=0$ with $u=t=m_\pi^2$ \cite{Guiasu:1979}, 
\begin{eqnarray}
%   \frac{\alpha A^i(s,t,u)}{2 m_\pi}&=&
&&\frac{\alpha}{2 m_\pi}\left(
   A^i(0,m_\pi^2,m_\pi^2)+ s \frac{d}{d s} A^i(0,m_\pi^2,m_\pi^2)\right)
   =\beta_1^i + \frac{s}{12}\beta_2^i ,\nonumber \\ 
&&-\alpha m_\pi\left( B^i(0,m_\pi^2,m_\pi^2)
   + s\frac{d}{d s} B^i(0,m_\pi^2,m_\pi^2) \right )
   =(\alpha_1+\beta_1)^i + \frac{s}{12}(\alpha_2+ \beta_2)^i,
\end{eqnarray}
where the superscripts $i=N,C$ denote the neutral and charged 
pions, respectively, and $\alpha\simeq 1/137$ is the fine structure constant. The Born 
term, arising in the case of the charged pion, is removed from the amplitudes. 

The NJL Lagrangian used in this work contains pseudoscalar isovector and 
scalar isoscalar four-quark interactions and is minimally coupled to the 
electromagnetic field.% Since we do not take into account explicit 
The diagrams of Fig.~\ref{fig:graphs} provide polarizabilities which scale 
as $N_c^0$. Besides the quark one-loop diagrams, it is expected that the pion 
loops yield important contributions, mainly in the case where the 
tree-level results are absent (in the NJL model the chiral counting of meson 
tree-level results are classified in Refs.~\cite{Bajc:1996,Bernard:1992}). 
We include the lowest 
model-independent pion-loop diagram at the $p^4$ order, calculated within 
$\chi$PT in Refs.~\cite{Bijnens:1988,Donoghue:1988,Donoghue:1993}, and known 
to be the only non-vanishing contribution to the amplitude $A$ at this order 
in the neutral channel. 
The pion loop in the charged mode contributes only to the quadrupole 
polarizabilities, with half 
the strength of the neutral quadrupole case. 
The pion loop as well as the $\sigma$-exchange diagram of Fig.~\ref{fig:graphs}
enter only the amplitude $A$. 

The amplitude $B$ for the 
neutral (charged) mode is completely determined by the quark box (quark box + pion exchange) diagrams,
starting from the $p^6$ order for the dipole and from the $p^8$ order for the quadrupole polarizabilities. 
Thus the combinations $(\alpha_j+\beta_j)^i$, $(j=1,2)$, to which the $B$ 
amplitude leads, provide a genuine test of the dynamical predictions of the 
NJL model at the leading order in $1/N_c$, as they are insensitive to the 
lowest-order $\chi$PT corrections. 
  
All quark one-loop integrals are regularized using the Pauli-Villars 
prescription with one regulator $\Lambda$ and two subtractions, which is consistent with the requirements of gauge invariance. 

In Table~\ref{param} we collect the model parameters, obtained by 
fitting the physical pion mass and the weak decay constant. The 
parameters of the model are the four-quark coupling constant $G$, the 
cutoff $\Lambda$, and the current quark mass $m$. These are determined by the choice of 
$f_\pi=93.1$~MeV, $m_\pi=139$~MeV (charged mode) or $m_\pi=136$~MeV 
(neutral mode), and the constituent quark mass, $M$. 
\begin{table}[tb]
\caption{The NJL model parameters for the charged and neutral channels, with 
the input marked by * and $f_\pi^*=93.1$~MeV.}
\label{param}
\begin{center}
\begin{tabular}{|c|c|c|c|c|}
\hline
    $M^*$ [MeV] & $m_\pi^*$ [MeV] &$m$ [MeV] & $G$ [GeV${}^{-2}$] 
    &$\Lambda$ [MeV] \\ \hline
%250 & 139 & 5.8 & 8.49 & 964 \\
%250 & 136 & 5.6 & 8.52 & 963 \\
300 & 139 & 7.5 & 13.1 & 827 \\
300 & 136 & 7.2 & 13.1 & 827 \\
%350 & 139 & 8.4 & 17.2 & 767 \\
%350 & 136 & 8.1 & 17.3 & 765 \\
\hline
\end{tabular}
\end{center}
\end{table}
In Table~\ref{tab:anat} we present the anatomy of our result for the case 
$M=300$~MeV. We display separately all the gauge invariant contributions 
to the polarizabilities: the  box (for neutral polarizabilities), box + pion 
exchange diagram (for the charged polarizabilities), the $\sigma$ exchange, 
and the pion loop.
The pion exchange diagram arises only for the 
charged channel and builds together with the box a gauge invariant amplitude.   
\begin{table}[tb]
\caption{The dipole (in units of $10^{-4} {\rm fm}^3$) and quadrupole (in units
of $10^{-4} {\rm fm}^5$) neutral pion polarizabilities. The first row 
shows our model prediction at $M=300$ MeV.} 
\label{tab:N}
\begin{center}
\begin{tabular}{|l|c|c|c|c|}
\hline
& $(\alpha_1+\beta_1)_{\pi^0}$ & $(\alpha_1-\beta_1)_{\pi^0}$ 
& $(\alpha_2+\beta_2)_{\pi^0}$ & $(\alpha_2-\beta_2)_{\pi^0}$ 
\\ \hline
%$M=250$~MeV & 1.13 & -2.05 & -0.33 & 46.3 
%\\
$M=300$~MeV & 0.73 & -1.56 & -0.14 & 36.1             
\\
%$M=350$~MeV & 0.50 & -1.32 & -0.07 & 30.6 
%\\
\hline 
DR fit \cite{Fil'kov:2005} &$0.98\pm 0.03$ &$-1.6\pm 2.2$ 
                               &$-0.181\pm 0.004$ &$39.70\pm 0.02$
\\
%\hline
DSR \cite{Fil'kov:2005}      &$0.802 \pm 0.035$ &$-3.49\pm 2.13$
                             &$-0.171\pm 0.067$ &$39.72\pm 8.01$ 
\\ 
$\chi$PT \cite{Bellucci:1994,Gasser:2005} 
    &$1.1\pm 0.3$ &$-1.9\pm 0.2$ &$0.037\pm 0.003$ &$37.6 \pm 3.3$ 
\\ 
\hline
\end{tabular}
\end{center}
\end{table}
\begin{table}[tb]
\caption{Same as in Table~\ref{tab:N} for the dipole and quadrupole charged 
pion polarizabilities.} 
\label{tab:V}
\begin{center}
\begin{tabular}{|l|c|c|c|c|}
\hline
          &$(\alpha_1+\beta_1)_{\pi^\pm}$ &$(\alpha_1-\beta_1)_{\pi^\pm}$ 
          &$(\alpha_2+\beta_2)_{\pi^\pm}$ &$(\alpha_2-\beta_2)_{\pi^\pm}$ 
\\ \hline
%$M=250$~MeV & 0.28                 & 10.4                 &  0.46                 &   23.8             \\
$M=300$~MeV & 0.19                 & 9.4                  &  0.20                 &   17.5              \\
%$M=350$~MeV & 0.13                 & 8.3                  &  0.10                 &   14.2             \\
\hline 
DR fit \cite{Fil'kov:2006} &$0.18^{+0.11}_{-0.02}$ &$13.0^{+2.6}_{-1.9}$  
                                  &$0.133\pm 0.015$     &$25.0^{+0.8}_{-0.3}$
\\ 
%\hline
DSR \cite{Fil'kov:2006} &$0.166\pm 0.024$ &$13.60\pm 2.15$ 
                        &$0.121\pm 0.064$ &$25.75\pm 7.03$ 
\\ 
$\chi$PT \cite{Bellucci:1994,Gasser:2006} &$0.16$  
                                          &$5.7\pm 1.0$  
                                          &$-0.001$  
                                          &$16.2$  
\\ 
\hline
\end{tabular}
\end{center}
\end{table}
\begin{table}[tb]
\caption{Contribution of various diagrams for  
$M=300$~MeV. Units as in Table \ref{tab:N}.} 
\label{tab:anat}
\begin{center}
\begin{tabular}{|l|c|c|c|c|}
\hline
& box + $\pi$-exchange & $\sigma$-exchange & pion-loop & total  \\ 
\hline  
$(\alpha_1+\beta_1)_{\pi^0}$   &   0.73 &  0    &  0     & 0.73   \\
$(\alpha_1-\beta_1)_{\pi^0}$   & -11.13 & 10.57 & -1.0   & -1.56  \\
$(\alpha_2+\beta_2)_{\pi^0}$   & -0.144 &  0    &  0     & -0.144 \\
$(\alpha_2-\beta_2)_{\pi^0}$   &  5.09  & 9.07  & 21.97  &  36.13 \\ 
\hline
$(\alpha_1+\beta_1)_{\pi^\pm}$ &  0.189 &  0     &   0    & 0.189  \\
$(\alpha_1-\beta_1)_{\pi^\pm}$ & -0.977 & 10.36  &   0    & 9.39   \\
$(\alpha_2+\beta_2)_{\pi^\pm}$ &  0.198 &  0     &   0    & 0.198  \\
$(\alpha_2-\beta_2)_{\pi^\pm}$ & -1.63  &  8.87  &  10.29 &  17.54 \\ 
\hline
\end{tabular}
\end{center}
\end{table}
Let us first comment on the channels involving the difference of the electric 
and magnetic polarizabilities: 

$\bullet$ $(\alpha_1-\beta_1)_{\pi^0}$: here the box 
contribution is largely canceled by the scalar exchange. At the $p^4$-order 
of the chiral counting they cancel exactly \cite{Bajc:1996}. The higher-order
contributions %are quark-mass dependent, decreasing as the constituent quark mass increases.
have a slow convergence rate, at $p^8$-order one reaches only 
about $50\%$ of the full sum. 

$\bullet$ $(\alpha_1-\beta_1)_{\pi^\pm}$: contrary to 
the neutral channel, the size of the $\sigma$-exchange diagram for this 
combination is about an order of magnitude larger than the box + pion 
exchange diagram, and it becomes the most important contribution. 
The pion loops are absent. 

$\bullet$ $(\alpha_2-\beta_2)_{\pi^\pm}$: the pattern 
observed for the charged dipole difference repeats itself for the quadrupole polarizabilities. However 
in this case the subleading in the $1/N_c$ counting pion-loop diagram has the 
same magnitude as the $\sigma$-exchange term. 

$\bullet$ $(\alpha_2-\beta_2)_{\pi^0}$: the pion loop dominates,  $\sim 2\times\sigma$-exchange.

$\bullet$ Finally, regarding the sums of polarizabilities, we
 stress that the values (including the overall signs) of  
$(\alpha_2+\beta_2)_{\pi^0,\pi^\pm}$ are totally determined by the 
gauge-invariant quark box or the box + pion exchange contribution. This is 
a key result of the presented
calculation. The sign is stable when the model parameters are changed. 
The magnitude depends on the value chosen for the constituent 
quark mass, but the best overall fit to the other empirical data, typically 
yielding $M\sim 300$~MeV, also yields the optimum values for the 
polarizabilities \cite{Hiller:2009}.  Moreover, the main part of the box contribution comes from the 
first non-vanishing $p^8$-order term in the chiral expansion.  
Based on this fact we expect that the contact term of the $p^8$ 3-loop 
calculation in $\chi$PT may also play an important role in reversing the 
signs of the 2-loop order results for these quantities.

\small{
 
}
\end{document}